\documentclass[10pt]{iopart}

\usepackage{graphicx}
\begin{document}

\title[Universal Crossover in Improved Perturbation Theory]{Universal Crossover in Perturbation Theory with a Large Field Cutoff}

\author{L. Li\dag\ and Y. Meurice\dag\ \ddag
\footnote[3]{To
whom correspondence should be addressed (yannick-meurice@uiowa.edu)}
}
\address{\dag\ Department of Physics and Astronomy\\ The University of Iowa\\
Iowa City, Iowa 52242 \\ USA }
\address{\ddag\ Also at the Obermann Center for Advanced Study, University of Iowa}

\date{\today}

\begin{abstract}
For $\lambda \phi^4$ models, the introduction of a large field cutoff
improves significantly the accuracy that can be reached with  
perturbative series but the calculation of the modified coefficients remains 
a challenging problem. We show that this problem can be solved 
numerically, and in the limits of large and small field cutoffs, for 
the ground state energy of the anharmonic oscillator. 
For the two lowest orders, the approximate formulas obtained in the large field 
cutoff limit extend unexpectedly far in the low field cutoff region.
For the higher orders, the transition between the small field cutoff regime 
and the large field cutoff regime can be described in terms of a universal 
function.
\end{abstract}
\section{Introduction}

Perturbative methods and Feynman diagrams have played an important role in the development 
of quantum field theory and its applications.  
However, perturbative series usually have a zero radius of convergence\cite{leguillou90}. 
For scalar models with $\lambda \phi^4$ interactions, the 
coefficients of perturbative series grow factorially. For any fixed, 
strictly positive, value of $\lambda$, there exists an order beyond which adding 
higher order terms diminishes the accuracy. 
This feature will restrict our ability to perform high precision tests of the standard model (for instance,  $g-2$ of leptons and the hadronic width of the $Z^0$) during the next decades.

The large order behavior of the
series is dominated by large field configurations which have little effect on low energy observables.
Introducing a large field cutoff \cite{pernice98,convpert} in the path integral formulation of 
scalar field theory, dramatically improves the large order behavior of the perturbative 
series. In two non-trivial examples \cite{convpert},  this procedure
yields series that apparently have finite radii of convergence and tend to values  
that are exponentially close to the exact ones. This also allows us to define the theory for 
negative or complex values of $\lambda$, a subject that has raised a lot of interest 
recently \cite{bender98,gluodyn04}. An important feature of this approach is that 
for a perturbative expansion at a given order in $\lambda$, it 
is apparently possible to determine 
an optimal field cut using the strong coupling expansion \cite{optim03,perfect05}, 
bridging the gap between the two expansions. In other words, the modified perturbative methods allows us to take into account non-perturbative effects. 

Despite these promising features, 
calculating the modified coefficients remains a challenging technical problem. 
While developing a new perturbative method (see e.g. Ref. \cite{buckley92,duncan92} for the $\delta$ expansion), it is customary to demonstrate the advantages of a method 
with simple integrals and the non-trivial, but well-studied\cite{bender69,leguillou90}, case of the anharmonic oscillator. A simple integral has been discussed in Ref. \cite{optim03}.
In this Letter, we show not only that this program can be completed in the 
case of the anharmonic oscillator, but also that the results show remarkable properties:
\begin{itemize}
\item
For the two lowest orders, the approximate formulas obtained in the large field 
cutoff limit extend unexpectedly far in the low field cutoff region.
\item 
For the higher orders, the transition between the small field cutoff regime 
and the large field cutoff regime can be described in terms of a universal 
function.
\end{itemize}

\begin{figure}\begin{center}  
\label{fig:num}               
\includegraphics[width=0.6\textwidth]{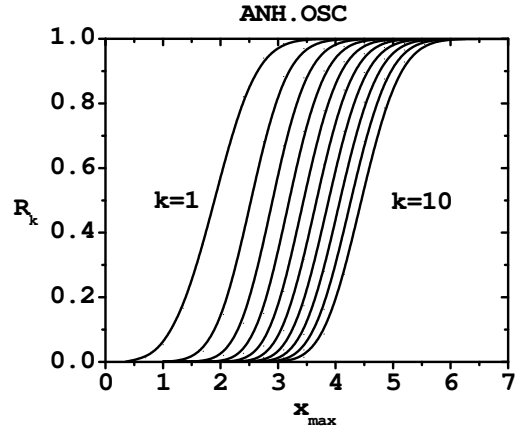}
\caption{The numerical values of $R_{k}(x _{\max
})
$ for $k$ going from 1 to 10.} \label{fig:an-ratio}
\end{center}
\end{figure}

In the following, we define the model considered, discuss the numerical calculation of the modified coefficients, 
the approximate methods at small and large field cutoff and the universal description of the crossover.
We use the quantum mechanical notations $x$ and $\omega$ instead 
of the field theoretical notations $\phi$ and $m$. 
In all numerical applications, we use $\omega=1$.
In the path integral formulation, 
introducing a field cutoff means that the paths with $|x(t)|>x_{max}$ for some $t$ 
are excluded. In the operator language, it means that the wave function $\Psi(x)$ should vanish for $|x|>x_{max}$. More specifically, we will consider 
the Hamiltonian 
\begin{equation}
H=\frac{p^{2}}{2}+V(x)\ ,
\end{equation}
with
\begin{equation}
V(x)=\left\{
\begin{array}{ccc}
\frac{1}{2}\omega ^{2}x^{2}+\lambda x^{4}\quad & {\rm if}& |x| < x_{\max}  \\
\infty \quad & {\rm if} & |x| \geq x_{\max}
\end{array}
\right.
\end{equation}
Our main interest if the calculation of the modified perturbative series for the 
ground state energy:
\begin{equation}
E_0(x_{max})=\omega \sum_{k=0}^{\infty }E_0^{(k)}(x_{max})(\lambda/\omega^3) ^{k}\ ,
\label{eq:eexp}
\end{equation}
Accurate numerical values for the $E_0^{(k)}(x_{max})$ can be obtained by 
combining the method of Ref. \cite{arbacc} with an expansion in $\lambda$. 
The technical aspects will be reported in a more detailed article. 
When $x_{max}\rightarrow\infty$, we recover the series that has been calculated to 
high order in Ref. \cite{bender69}. In order to allow comparison among the different orders, we define the ratios
\begin{equation}
R_k(x_{max})\equiv E_0^{(k)}(x_{max})/E_0^{(k)}(\infty)\ ,
\end{equation}
which all tend to 1 in the 
$x_{max}\rightarrow\infty$ limit. The numerical values of these ratios are shown in 
Fig. \ref{fig:num}. A striking feature is that the curves for the various orders have approximately the same shape.

The asymptotic 
behavior of the series can be studied from this data. If $x_{max}$ is small enough 
(say below 2 in Fig. \ref{fig:num}), then most of the coefficients have $R_k\sim 0$ 
(asymptotic regime)  
and a linear behavior is observed for ${\rm ln}(|E^{(k)})$ when $k$ increases.
This allows us to infer that the series converges for $|\lambda|<Cx_{max}^{-6}$ with 
$C\simeq 65$.

\begin{figure}\begin{center}                                
\includegraphics[width=0.7\textwidth]{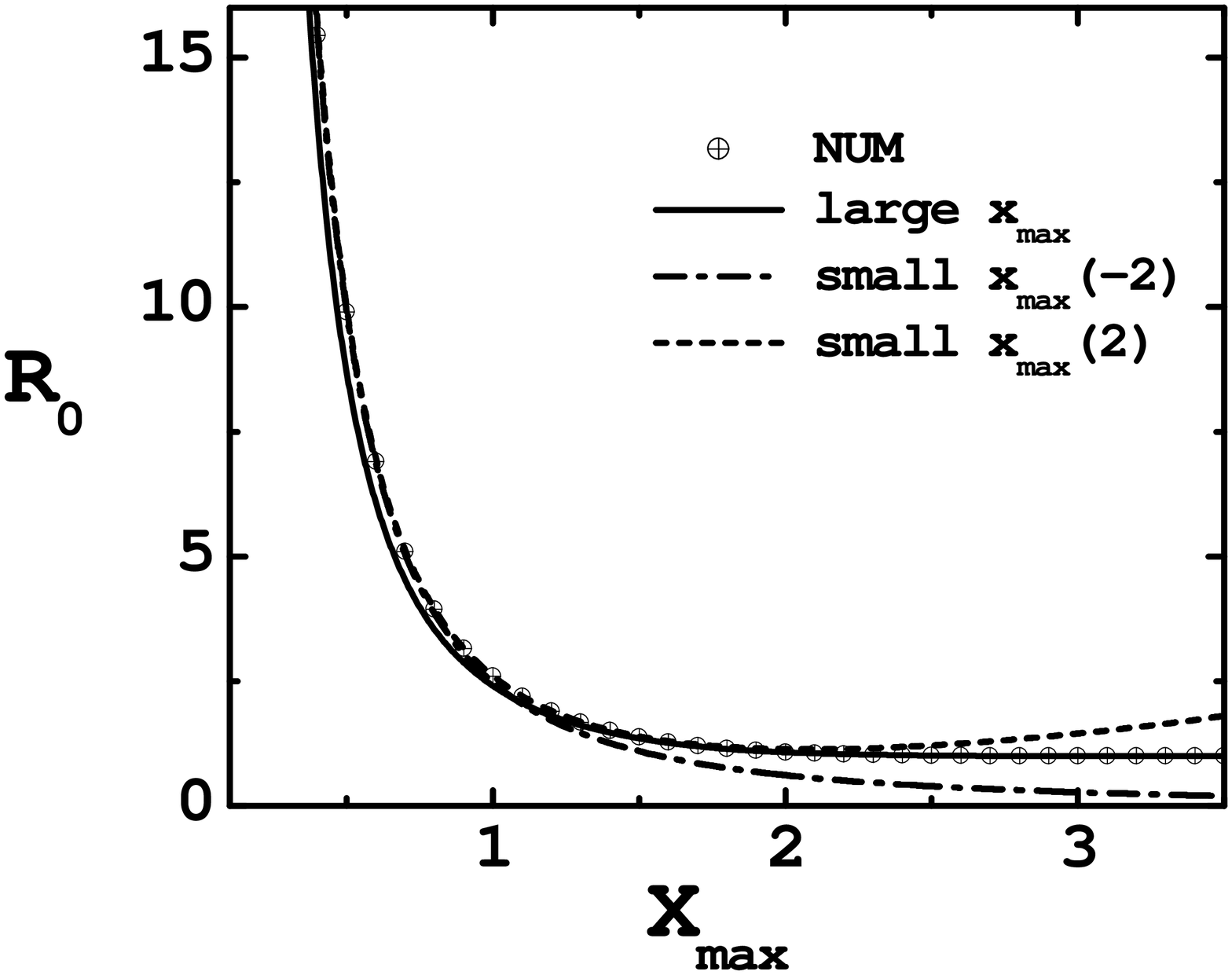}
\includegraphics[width=0.7\textwidth]
{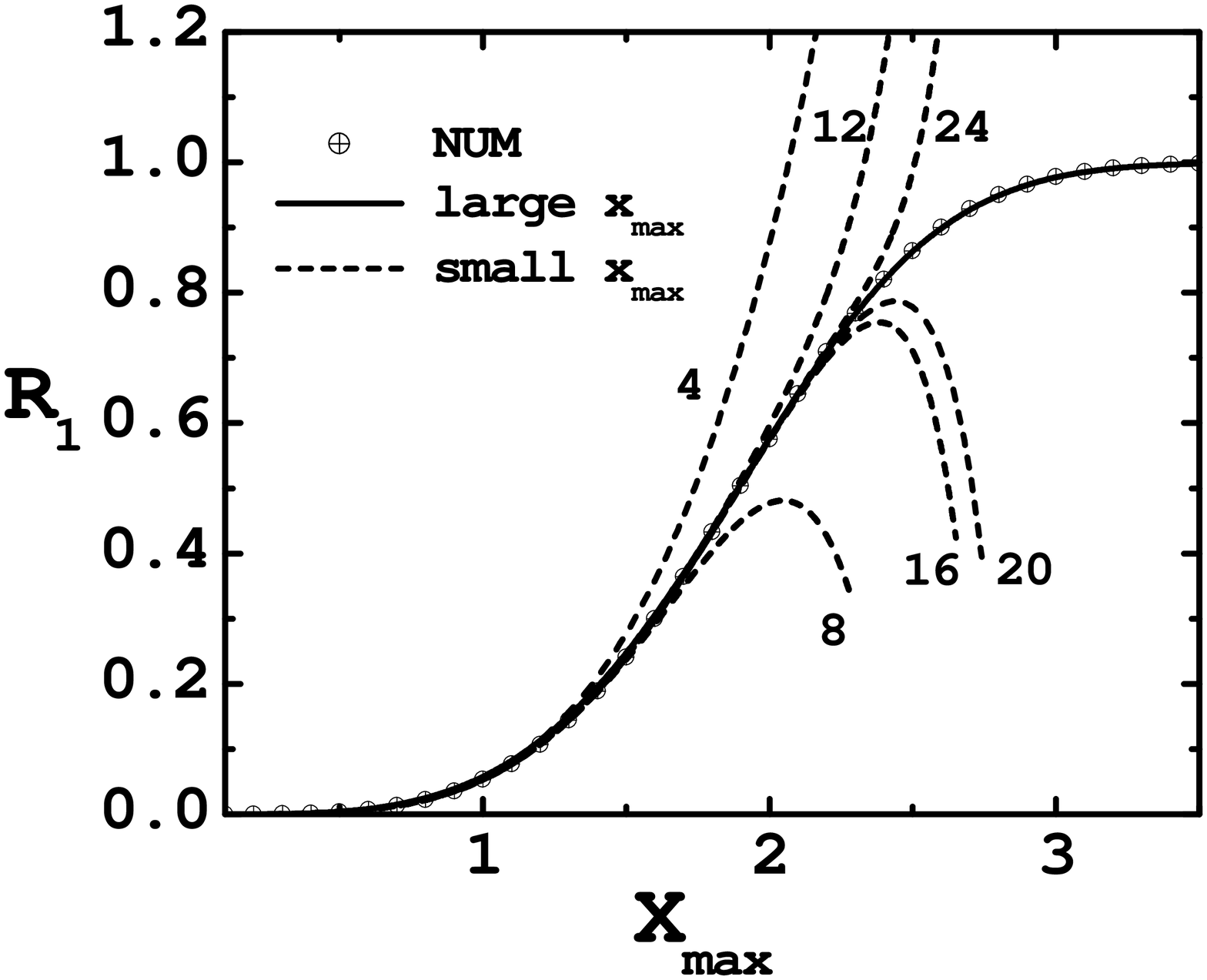}
\caption{Numerical values of $R_0(x_{max})$ and $R_1(x_{max})$ . The solid lines 
represent the large $x_{max}$ expressions. The broken lines represent  lowest orders in the small $x_{max}$ approximation.}
\label{fig:r1}
\end{center}\end{figure}

We now discuss an approximate method of calculation of 
$1-R_{k}(x _{\max
})$ in the large $x_{max}$ limit. For $k=0$, our expansion parameter will be
$\epsilon=(E_0^{(0)}-1/2)$ in terms of which $R_0=1+2\epsilon$. Assuming a ground state wave function of the 
form 
\begin{equation}
\Psi_0^{(0)}(x)\propto {\rm e}^{-\omega x^2/2}\ (1+\epsilon G(x)+\dots)
\end{equation}
and expanding at first order in $\epsilon$, 
we obtain a second order, inhomogeneous, linear differential equation for $G$. 
Remarkably, this equation can be integrated exactly and 
\begin{equation}
	G(x)=-2\omega \int_{0}^x dy {\rm e}^{\omega y^2}\int_{0}^y dz {\rm e}^{-\omega z^2}\ ,
	\label{eq:g}
\end{equation}
From the boundary condition $\Psi_0^{(0)}(x_{max})=0$ we obtain
$\epsilon=-1/G(x_{max})$. The asymptotic 
expression agrees with the semi-classical estimate of Ref. \cite{convpert}.
We can then use our first order result to calculate
\begin{equation}
	E^{(1)}_0=2\omega^2\int_0^{x_{max}}|\Psi_0^{(0)}(x)|^2x^4 \ .
\end{equation}
The comparison with numerical data is shown in Fig. \ref{fig:r1}.
Unexpectedly, the approximation stays accurate in regions where $\epsilon$ is by no 
means small. Integral formulas such as Eq. (\ref{eq:g}) for 
the excited states will be discussed 
in a more detailed publication.
For higher orders, there is solid numerical evidence for a large $x_{max}$ behavior of the form
\begin{equation}
1-R_k(x_{max})
\propto
x_{\max }^{4k+1}e^{-\omega x_{\max }^{2}}\ .
\end{equation}

We now turn our attention to the opposite limit of small $x_{max}$. 
The entire potential (including the quadratic term) can then be treated as a 
perturbation. At zero-th order, we have the solvable problem of a free particle in a box of size $2x_{max}$. Using the conventional techniques of nondegenerate time-independent 
perturbation theory, or numerical method (after a rescaling that allows the 
use a numerical value for the rescaled $x_{max}$), we can calculate the 
coefficients of the series: 
\begin{equation}
E_0^{(k)}=\sum_{m=0}^{\infty}E_n^{(k,m)}(\omega x_{max}^2)^{3k+2m-1}\ .
\end{equation}
The lowest order approximations for $k=$ 0 and 1 are shown in Fig. \ref{fig:r1}. Numerical calculations of 6 terms in the series indicate that 
the series have a finite radius of convergence that seem to shrink slowly when the 
order increases. For $k=0$ ($k=5$), we apparently have convergence for $\sqrt{\omega}x_{max}<3.3$ ($\sqrt{\omega}x_{max}<2.3$). In the case $k=0$ or 1, this
is sufficient to have a good overlap between the validity range of the two limits.
However, for higher orders, other methods are necessary to deal with the crossover 
between the two regimes.

The shape similarities 
observed in Fig. \ref{fig:num} suggest to parametrize $R_k$ in terms of a unique, 
universal,  
function $U$, that can be shifted by a $k$-dependent quantity that we denote 
$x_0(k)$:
\begin{equation}
	R_k(x_{max})
	\simeq U(x_{max}-x_0(k))\ .
\end{equation}
In order to determine $x_0$, we used the derivative of $R_k$ with respect to $x_{max}$ that is bell shaped and we parametrized it as $B_k e^{-A_k(x-x_{0}(k))^{2}}$. Numerical fits and the generic form of the 
terms of the perturbative series at large $x_{max}$, led us to infer 
that $A=2\omega$. Adding exponentially small tails on the left (for $k$ 
large enough) and imposing the correct matching on the right, we obtained: 
\begin{equation}
U(x_{max})=(2\omega/\pi)^{1/2}	\int_{-\infty }^{x_{\max }}dx\exp(-2\omega x^{2} )\ .
\label{eq:u}
\end{equation}
The dependence on the order is encoded in the shift 
$x_0(k)$. Empirically, they can be fitted quite well with 
\begin{equation}
	x_0(k)\simeq 0.87 + 1.13\sqrt{k}\ .
\end{equation}

\begin{figure}\begin{center}                               
\includegraphics[width=0.7\textwidth]{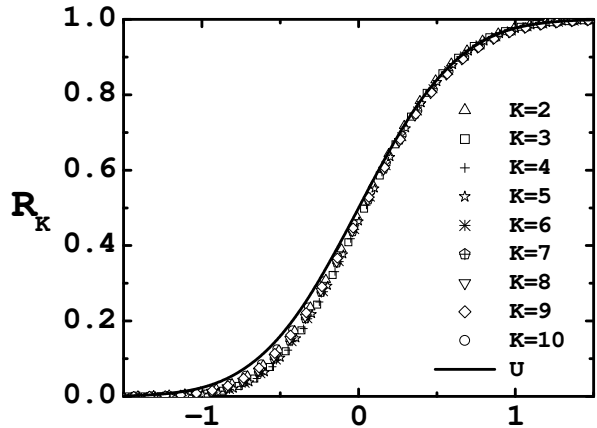}
\caption{ $R_k(x_+x_0(k))$ for $k=2, \dots 10$ and the universal function 
$U(x)$ of Eq. (\ref{eq:u}).}
\label{fig:coll}
\end{center}\end{figure}

The $R_k$ shifted by $x_0(k)$ are shown in Fig. \ref{fig:coll}. 
One can see that the collapse of the various curves is quite good. The function 
$U$ fits the collapsed data quite well on the upper right, but appears to be slightly above the data on the lower left. This can be explained by the 
fact that the function satisfies the duality relation which exchanges the small and large field cutoff regions
\begin{equation}
	U(-x)=1-U(x)\ ,
	\label{eq:dual}
\end{equation}
which cannot be exact since the approach of 0 when $x_{max}\rightarrow 0$ is power like 
while in the large $x_{max}$ limit, the approach of 1 is exponentially small.
In these two limits, $U$ fails to provide the correct behavior and we need to 
resort to the approximations discussed before.

We expect similar features in higher dimensional scalar field theory with a UV 
regulator. Generically, a modified perturbative calculation involves a few 
coefficients with values close to their usual ones, a few coefficients in the crossover region, and the rest of the coefficients taking values much smaller than 
in regular perturbation theory. These three regimes are reminiscent of the three regimes encountered when calculating renormalization group flows between two fixed points. One should notice the similarities between the graphs presented here and those of Refs. \cite{bagnuls01,pelissetto98} where this type of crossover behavior is studied.
In the same way, the approximate duality of Eq. (\ref{eq:dual}) reminds of the 
duality between fixed points of a simplified renormalization group equation \cite{dual} or the exchange between the perturbative and non-perturbative sectors 
of quasi-exactly solvable periodic potentials
\cite{dunne02}. We plan to make these analogies more specific  
(or disprove these ideas), by 
making explicit calculations with Dyson's hierarchical model. 
Namely, we plan to compare the relative weighs of the three types of 
perturbative terms discussed above with the relative increases
in the three regions of the renormalization group flows discussed 
above for the zero-momentum $n$-point functions.

This research was supported in part by the Department of Energy
under Contract No. FG02-91ER40664. We thank J. Cook and B. Kessler for 
valuable discussions.


\end{document}